\pdfoutput=1

\documentclass[11pt]{article}

\usepackage[preprint]{acl}

\usepackage{times}
\usepackage{latexsym}

\usepackage[T1]{fontenc}

\usepackage[utf8]{inputenc}

\usepackage{microtype}
\usepackage{graphicx}
\usepackage{float}
\usepackage{amsmath}
\usepackage{diagbox}
\usepackage{multirow}
\usepackage{array}
\usepackage{titlesec}
\usepackage{pifont}
\usepackage[normalem]{ulem}
\usepackage{enumitem}
\usepackage{subfig}
\usepackage{hhline} 
\usepackage{amsfonts}
\usepackage{parskip} 

\usepackage{array}
\newcommand{\PreserveBackslash}[1]{\let\temp=\\#1\let\\=\temp}
\newcolumntype{C}[1]{>{\PreserveBackslash\centering}p{#1}}
\newcolumntype{R}[1]{>{\PreserveBackslash\raggedleft}p{#1}}
\newcolumntype{L}[1]{>{\PreserveBackslash\raggedright}p{#1}}

\usepackage{inconsolata}

\title{DAPR: A Benchmark on Document-Aware Passage Retrieval}

\author{
Kexin Wang\textsuperscript{\textnormal{1}}, 
Nils Reimers\textsuperscript{\textnormal{2}},
Iryna Gurevych\textsuperscript{\textnormal{1}}\\
 \textsuperscript{\textnormal{1}} Ubiquitous Knowledge Processing Lab (UKP Lab)\\
Department of Computer Science and Hessian Center for AI (hessian.AI)\\
Technical University of Darmstadt  \\
 \textsuperscript{\textnormal{2}} Cohere\\
 \url{www.ukp.tu-darmstadt.de}
}

\begin{document}
\maketitle
\begin{abstract}
The work of neural retrieval so far focuses on ranking short texts and is challenged with long documents. There are many cases where the users want to find a relevant passage within a long document from a huge corpus, e.g. Wikipedia articles, research papers, etc. We propose and name this task \emph{Document-Aware Passage Retrieval} (DAPR). While analyzing the errors of the State-of-The-Art (SoTA) passage retrievers, we find the major errors (53.5\%) are due to missing document context. This drives us to build a benchmark for this task including multiple datasets from heterogeneous domains. In the experiments, we extend the SoTA passage retrievers with document context via (1) hybrid retrieval with BM25 and (2) contextualized passage representations, which inform the passage representation with document context. We find despite that hybrid retrieval performs the strongest on the mixture of the easy and the hard queries, it completely fails on the hard queries that require document-context understanding. On the other hand, contextualized passage representations (e.g. prepending document titles) achieve good improvement on these hard queries, but overall they also perform rather poorly. Our created benchmark enables future research on developing and comparing retrieval systems for the new task. The code and the data are available\footnote{\url{https://github.com/UKPLab/acl2024-dapr}}.
\end{abstract} 

\section{Introduction}
\begin{figure}[t]
  \centering
  \includegraphics[width=60mm]{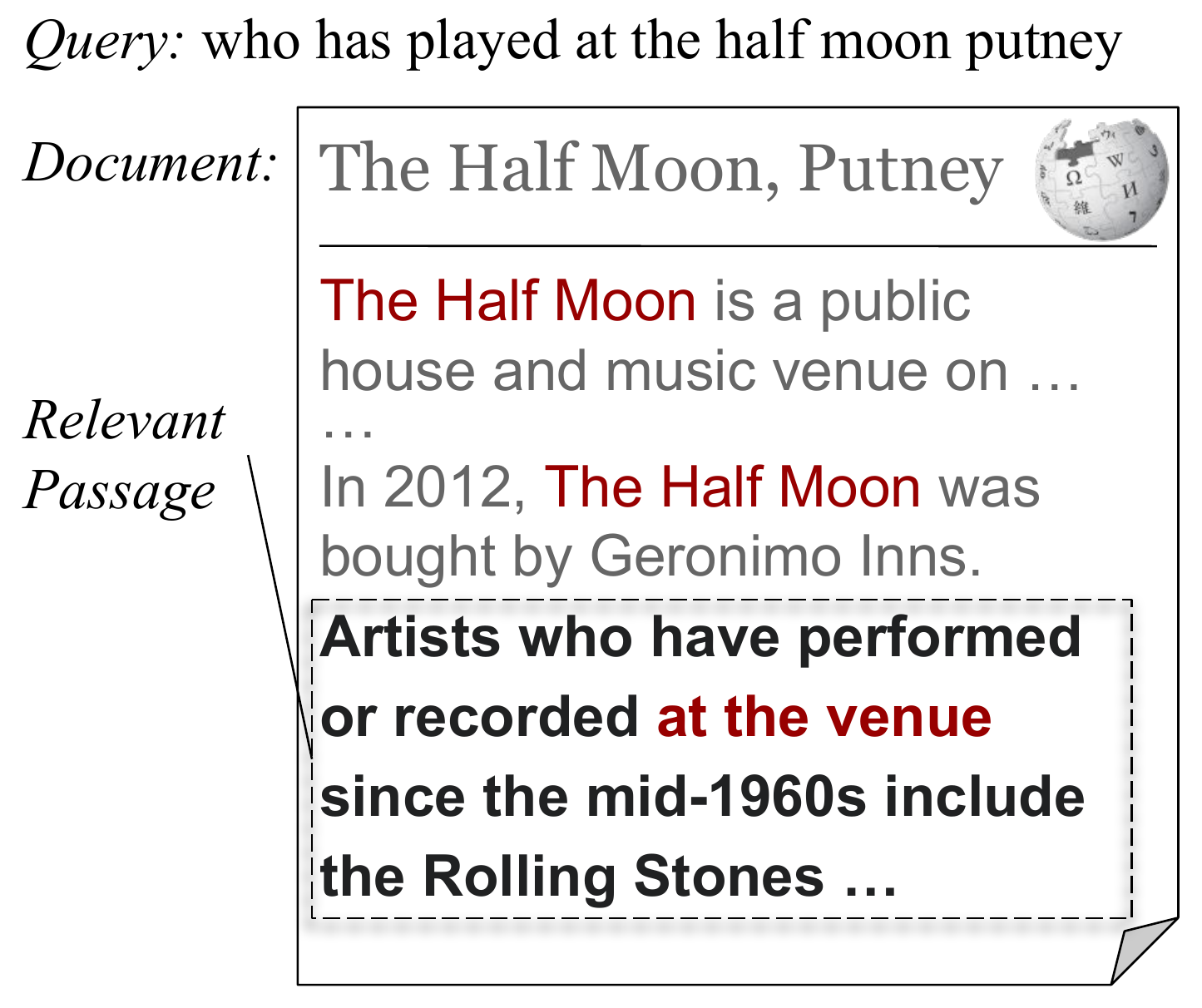}
  \caption{An example instance from DAPR. To find the relevant passage to the query, the retriever needs to utilize the document context, which in this case means coreference resolution for the noun \emph{the venue}. See other categories of the document context and examples in~\autoref{sec:nq_hard}.
  }
  \label{fig:motivative-example}
\end{figure}
Information Retrieval (IR) helps efficiently locate relevant information from a vast resource collection, acting as a central component of many natural language applications. Traditional approaches like BM25 compute simple statistics such as the frequency of the matched terms~\cite{bm25}. Recent approaches apply neural networks to represent queries and passages into vector representations, extending the task modeling from simple term matching to complex semantic matching achieving better effectiveness~\cite{xiao-etal-2022-retromae,spladev2,santhanam-etal-2022-colbertv2}.

Despite their success, these neural approaches are usually limited to short passage inputs, e.g. 512 tokens due to expensive operations such as self-attention~\cite{transformer,devlin-etal-2019-bert} in their architectures. Such short-passage retrieval faces severe challenges in real-world scenarios, where long documents such as Wikipedia\footnote{\url{https://www.wikipedia.org/}} articles, scientific papers, etc. can easily go beyond this length limit. Recent work proposes new memory-efficient architectures to encode much longer document inputs~\cite{dai-etal-2019-transformer,longformer} and fulfill document-retrieval tasks~\cite{sedr}. However, returning a long document is still inefficient for a user to locate useful information. For example, \citet{kwiatkowski-etal-2019-natural} collects user queries from Google Search\footnote{\url{https://www.google.com/}} logs and annotates the relevant passage in Wikipedia pages. We find for 35.8\% of the queries, their relevant passage is located at the 7.6th paragraph on average (standard deviation 12.7), indicating a large further-search range. This adds tremendous extra effort for the users for information seeking.

To understand the ability of the retrieval systems for filling this gap, we propose the \emph{Document-Aware Passage Retrieval} (DAPR) task, where the retriever is required to consider the associated document context for returning relevant passages. An example is shown in~\autoref{fig:motivative-example}. In this case, the user asks for musicians that have played at a specific venue. However, the relevant passage does not mention the venue name but only the noun reference and the retriever needs to understand such document context for finding the correct passage. To gain insight into the challenges, we first carry out an error analysis for the SoTA passage retrievers (DRAGON+~\cite{lin-etal-2023-train}, SPLADEv2~\cite{spladv2distil}, ColBERRTv2~\cite{santhanam-etal-2022-colbertv2}) and BM25. We find the major errors (53.5\%) are due to missing document context, where the correct passage misses coreference resolution, the information of the underlying main topic, etc. In these cases, understanding the document context is necessary for relating the query to the correct passage. This motivates us to create a benchmark for this task including 5 datasets from heterogeneous domains that provide such annotations of the relevant passage within its associated document. 

In experiments, we test the approaches that extend the SoTA neural passage retrievers by introducing the document context to them in two types of approaches: (1) hybrid retrieval with BM25 and (2) contextualized passage representations, which inform the passage representations with the document context. We find while the hybrid-retrieval systems achieve the strongest performance on the mixture of the easy queries and the hard queries, they fail to process the latter case where understanding the document context is necessary. Contextualized passage representations, on the other hand, can achieve good improvement on these hard queries, but overall perform rather poorly. This presents new exciting research opportunities for developing improved retrieval methods that understand the context of documents during passage retrieval. The benchmark we developed enables the research community to develop, evaluate, and compare the retrieval systems on the new task.
\begin{table*}[t]
\centering
\resizebox{16cm}{!}{
\begin{tabular}{|l|l|c|c|c|c|c|c|c|c|c|c|} 
\hline
\multirow{2}{*}{\textbf{Name}} & \multirow{2}{*}{\textbf{Domain}} & \multirow{2}{*}{\textbf{\#Docs.}} & \multirow{2}{*}{\begin{tabular}[c]{@{}c@{}}\textbf{\#Psg.}\\\textbf{per doc.}\end{tabular}} & \multirow{2}{*}{\begin{tabular}[c]{@{}c@{}}\textbf{Psg.}\\\textbf{len.}\end{tabular}} & \multirow{2}{*}{\begin{tabular}[c]{@{}c@{}}\textbf{Query.}\\\textbf{len.}\end{tabular}} & \multirow{2}{*}{\begin{tabular}[c]{@{}c@{}}\textbf{Title}\\\textbf{len.}\end{tabular}} & \multirow{2}{*}{\begin{tabular}[c]{@{}c@{}}\textbf{Rel.~}\\\textbf{scales}\end{tabular}} & \multirow{2}{*}{\begin{tabular}[c]{@{}c@{}}\textbf{Depth of }\\\textbf{rel. passage}\end{tabular}} & \multicolumn{3}{c|}{\textbf{\#Queries / \#judgements per query}}  \\ 
\cline{10-12}
                               &                                  &                                   &                                                                                             &          &&                                                                             &                                                                                          &                                                                                                    & \textbf{Train} & \textbf{Dev} & \textbf{Test}                     \\ 
\hline
MS MARCO                         & Misc.                            & 1,359,163                         & –                                                                                           & 63.9  &7.2& 7.7                                                                    & Binary                                                                                   & –                                                                                                  & 92,905/1.0     & 2,839/1.0    & 2,722/1.0                         \\
Natural Questions                            & Wiki.                            & 108,626                           & 24.7                                                                                        & 105.5 &9.9& 4.7                                                                       & Binary                                                                                   & 7.6$\pm$12.7                                                                                       & 93,275/1.0     & 3,610/1.0    & 3,610/1.2                         \\
MIRACL                         & Wiki.                            & 5,758,285                         & 5.7                                                                                         & 105.3 &9.1& 5.0                                                                     & Binary                                                                                   & 13.4$\pm$21.3                                                                                      & 2,064/2.8      & 799/2.7      & 799/2.9                           \\
Genomics                       & Biomed.                          & 162,259                           & 77.9                                                                                        & 150.5 &14.0& 22.0                                                       & 3-scale                                                                                  & 38.0$\pm$38.6                                                                                      & –              & –            & 62/121.9                          \\
ConditionalQA                  & Gov.                             & 652                               & 106.1                                                                                       & 15.8  &58.7& 6.2                                                           & Binary                                                                                   & 38.8$\pm$36.9                                                                                      & 1,975/4.1      & 271/4.0      & 271/4.3                           \\
\hline
\end{tabular}}
\caption{Statistics of the datasets in DAPR. Depth of rel. passage indicates the position (starting from 1) of the relevant passage in its associated document. The distribution plots of this value are available in~\autoref{fig:depth_dists}. For MS MARCO, there exists no available gold paragraph segmentation.}
\label{tbl:stats}
\end{table*}

\section{Related Work}
\begin{description}[wide,itemindent=\labelsep]
    \item[Document Question Answering (DocQA)] is a similar task to DAPR, requiring the model to answer a question about an input document~\cite{DocQA,qasper,sun-etal-2022-conditionalqa}. Besides the general difference between QA and IR, the main distinction is that DocQA usually assumes the relevant document is given before asking a question. In DAPR, we expect the model to find the relevant information from a (large) collection of documents instead, as the user might not know which document should be referred to before inputting the query. Also due to this assumption, the questions in DocQA tasks are usually contextualized wrt. the given document, e.g. "which retrieval system was used for the baselines?" in QASPER~\cite{qasper}. In that case, the DocQA task cannot be converted into a DAPR task, because the questions are too general and unanswerable without knowing the referred document.

    \item[Long-document retrieval] One simple strategy to extend the passage retriever is taking the maximum of the passage relevance within its belonging document as the document relevance (named MaxP); or encoding only the first passage of the document (named FirstP)~\cite{ance}. This strategy is sub-optimal due to still ignoring document context. \citet{xiong-etal-2022-simple} compares different long-range attention modules on a document-retrieval task for Transformer-based neural networks~\cite{transformer}. \citet{sedr} propose a hierarchical neural network and show it can outperform the MaxP-based approaches. All these previous works do not study how to retrieve passages while considering their document context in contrast to DAPR.
    \item[Hybrid retrieval] involves multiple retrieval systems (usually BM25 and a neural retriever) for each single query. Among them, rank fusion~\cite{pinecone-fusion} combines the individual rankings from different systems into one via e.g. convex combination~\cite{dense-retrievers-require-interpolation}, reciprocal ranks~\cite{rrf}, etc. Previous work mainly studies rank fusion between passage rankings and we in this work examine the effectiveness of combining document ranking and passage ranking. As another typical way, hierarchical retrieval first retrieves documents and then retrieves passages from these documents~\cite{liu-etal-2021-dense-hierarchical,arivazhagan-etal-2023-hybrid}. Previous work claims hierarchical retrieval is effective, but we find its advantage is limited to only better retrieving self-contained passages where the passages themselves are enough for responding to the query.
    \item[Relation to pre-training tasks] One relevant line of research is about utilizing the document context as the training signal~\cite{ein-dor-etal-2018-learning,giorgi-etal-2021-declutr,Wu_Ma_Lin_Lin_Wang_Hu_2023}. For example, \citet{Wu_Ma_Lin_Lin_Wang_Hu_2023} proposes a new pre-training method that forces the encoder to compress one span well for the good reconstruction of another span from the same document. As a pre-training method, these approaches discard document context during inference while evaluating with the traditional passage-ranking/sentence-embedding tasks by looking at only the independent passages/sentences. On the other hand, our proposed new task, DAPR requires the retrieval system to model the document context during inference for a good evaluation score. As the emphasis of this work, we fill the gap by formalizing the problem in a well-defined way, which enables investigation and refinement of document-context approaches.
    
\end{description}

\section{The DAPR task and the Benchmark}
The DAPR task asks the systems to retrieve and rank the relevant passages given their associated (long) document. 

Formally, given a collection of passages $C=\{p_i\}_{i=1}^N$ and their associated documents $D=\{d_i\}_{i=1}^N$, for a query $q\in Q$, the retrieval system $s:Q\times C\times D\longrightarrow \mathbb{R} $ is required to return the top-$K$ passages $R=\{p_1, p_2, ... p_K\}$ st. $\forall j, s(q, p_j, d_j) \le \min_{p_i\in R}s(q, p_i, d_i)$.


\subsection{NQ-Hard: The Hard Cases in Natural Questions}
\label{sec:nq_hard}

\begin{table*}[t]
\centering
\resizebox{15.5cm}{!}{
\small
\begin{tabular}{|m{1.5cm}|m{1cm}|m{2cm}|m{5cm}|m{8cm}|} 
\hline
\textbf{Category}      & \textbf{\%}                         & \textbf{Query}                                                                            & \textbf{Gold-relevant passage}      & \textbf{Retrieved (top-1)}  \\ 
\hline
Self-contained (excluded from NQ-hard)         & 39.0\%                    & how did early humans make use of stones during the prehistoric period            & \textbf{Prehistoric technology is technology that predates recorded history ... About 2.5 million years before writing was developed, technology began with the earliest hominids who used stone tools, which they may have used to start fires, hunt, and bury their dead. }                                                                                                                                                                                                                                                                                                                                                                                                                                                                                        & \ding{55} The Stone Age is a broad prehistoric period during which stone was widely used in the manufacture of implements with a sharp edge, a point, or a percussion surface. The period lasted roughly 2.5 million years, from the time of early hominids to Homo sapiens in the later Pleistocene era, and largely ended between 6000 and 2000 BCE with the advent of metalworking.[citation needed]                                                                                                                                                                                                                                                                                                                                                                                                                                                                                                                                                                                                                                                            \\ 
\hhline{=====}
\multicolumn{5}{|l|}{\textit{Categories in NQ-hard}}\\
\hline
Coreference resolution & 22.1\%                    & meaning of joy to the world by three dog night                                   & \begin{tabular}[c]{@{}m{5cm}@{}}\textit{Title:~Joy to the World (Three Dog Night song)}\\\textit{\textcolor{red}{"Joy to the World" is a song} written by Hoyt Axton and made famous by the band Three Dog Night. ...}\\\textbf{\textcolor{red}{The song}, which has been described by members of Three Dog Night as a "kid\textbackslash{}'s song" and a "silly song" ...
}\end{tabular}    

& \ding{55} "Joy to the World" is a song written by Hoyt Axton, and made famous by the band Three Dog Night. The song is also popularly known by its opening lyric, "Jeremiah was a bullfrog". Three Dog Night originally released the song on their fourth studio album, Naturally in November 1970 and subsequently released an edited version of the song as a single in February 1971.[1]                                                                                                                                                                                                                                                                                                                                                                                                                                                                                                                                                                                                                                                                        \\ 
\hline
Main topic             & 21.3\%                    & 

who gave a speech to the democratic national convention in 1984 

& \begin{tabular}[c]{@{}m{5cm}@{}}\textit{\textcolor{red}{Title:~1984 Democratic National Convention}}

...\\\textbf{New York Governor Mario Cuomo gave a well-received keynote speech. Mondale's major rivals for the presidential nomination, Senator Gary Hart and Rev. Jesse Jackson, also gave speeches.}\end{tabular}

& \ding{51} In United States politics, at the 1984 Democratic National Convention, the keynote speaker Gov. Mario Cuomo of New York delivered a scathing criticism of ...                                                                                                                                                                                                                                                                                                                                                                                                                                                                                                                                                                                                                                                  \\ 
\hline
Multi-hop reasoning       & 10.0\%                        & who was \textcolor{red}{tammy} from \textcolor{red}{basketball wives} married to & 
\begin{tabular}[c]{@{}m{5cm}@{}}
\textit{Title:~\textcolor{red}{Tami Roman}}\\\textit{Tamisha Akbar Youngblood (born April 17, 1970), known ... she became one of the breakout stars of VH1's \textcolor{red}{Basketball Wives}.[2]}...\\\textbf{\textcolor{red}{Roman} was once married to NBA basketball player Kenny Anderson.[3] The marriage produced 2 children.[4]'}
\end{tabular} & \ding{55} Tammy Swanson (Patricia Clarkson) is the first ex-wife of Ron Swanson. Ron has been married to two different women, both named Tammy, and he hates and fears both of them ...  \\ 
\hline
Acronym                 & 1.2\%       & who has been chosen as the brand ambassador of the campaign 'beti bachao-beti padhao & \begin{tabular}[c]{@{}m{5cm}@{}}

\textit{\textcolor[rgb]{0.125,0.129,0.133}{The }\textcolor{red}{Beti Bachao, Beti Padhao (BBBP) }\textcolor[rgb]{0.125,0.129,0.133}{ scheme was launched on 22 January 2015 by Modi.. It aims ...}}\\\textbf{In 26 August 2016, Olympics 2016 bronze medallist Sakshi Malik was made brand ambassador for BBBP.[7]}\end{tabular}

& \ding{51} Currently, she is appointed as the brand ambassador for Beti Bachao, Beti Padhao programme by Telangana State Government.\\
\hline
\end{tabular}}
\caption{Categories of the errors in Natural Questions. The total number of query-passage pairs is 1,031. 66 (6.3\%) pairs are excluded due to the wrong annotation (i.e. not an answer) from the original data. Gold-relevant passages are \textbf{bold} and other related document context is \textit{italic}. The hints of the document context are in \textcolor{red}{red}. \ding{51} and \ding{55} indicate that the retrieved passage by ColBERTv2 is actually relevant or irrelevant under manual check. Note that due to the insufficient annotation of the original datasets (known as ``shallow pooling'' in~\citet{arabzadeh2022shallow,survivorship_bias}), there are cases where the retrieved passage is actually relevant but not labeled.
}
\label{tbl:nq_hard}
\end{table*}

To understand the importance of the DAPR task, we carry out an error analysis on Natural Questions for the SoTA passage retrievers (DRAGON+, SPLADEv2, and ColBERTv2) and BM25. We focus on the error cases where all of these retrievers achieve zero nDCG@10, which yields 895 queries\footnote{Corresponding to 1,031 query-passage pairs. As shown in~\autoref{tbl:stats}, each query corresponds to 1.2 gold relevant passages on average in the test split of NQ.} out of the total set of 3,610 queries. Then we manually check the query-passage pairs one by one and label them as self-contained or not. Here, self-contained pairs represent the examples where the passage itself is enough for responding to the query without additional information. This filtering process ends up with 479 queries\footnote{Corresponding to 516 query-passage pairs.} whose relevant passages are all not self-contained. This means 53.5\% of the error cases are due to out of passage context, where understanding document context is necessary to retrieve the passages. This drives us to create a benchmark for the DAPR task to evaluate and compare the retrieval systems.

We name these hard data instances in the original Natural Questions as \emph{NQ-hard}. In NQ-hard, we find the query-passage pairs can be grouped into four categories: (1)~\textbf{Coreference Resolution (CR):} key coreference information with the gold-relevant passage needs to be resolved by certain document context; (2)~\textbf{Main Topic (MT):} the gold-relevant passage can only answer the query by knowing the background topic (usually the title) of the document; (3)~\textbf{Multi-Hop Reasoning (MHR):} the inference path which connects the entities in the query and the gold-relevant passage includes other nodes in the document context; (4)~\textbf{ACronym (AC):} there appears in the relevant passage (or the query) an acronym which corresponds to the full name in the query (or the relevant passage) and the document context explains such a mapping. During data annotation, we manually assign these category labels to the query-passage pairs\footnote{Each pair can have multiple labels.}. The annotation results are shown in~\autoref{tbl:nq_hard}. 

\subsection{DAPR Datasets}
\label{sec:dataset}

Driven by the error analysis in~\autoref{sec:nq_hard}, according to the definition of the DAPR task above, we select five datasets and unify their formats to compose DAPR: MS MARCO~\cite{msmarco}, Natural Questions~\cite{kwiatkowski-etal-2019-natural}, MIRACL~\cite{miracl}, Genomics~\cite{genomics2006,genomics2007}, and ConditionalQA~\cite{sun-etal-2022-conditionalqa}. These datasets are from heterogeneous domains, making the evaluation robust. The details about these datasets and the preprocessing are available in~\autoref{sec:dataset_details}. For MS MARCO and Natural Questions, we view the original dev split as the test split and hold a sampled set of training examples\footnote{For MS MARCO, we sample examples from the QA task, which are not included in the training split of its original Passage Ranking task.} as our dev split. For the other datasets, we keep their original data splits. For the datasets except MS MARCO, the paragraph segments\footnote{In such cases, all the paragraph segments together form the same set of the passage collection $C$.} are available; for MS MARCO, such information does not exist and only the passage spans within the associated document in $C$ are available. The statistics are shown in~\autoref{tbl:stats}. The examples of the data instances are shown in~\autoref{tbl:examples}. 

\subsection{Evaluation}
Since the judgment can be multi-scale in DAPR, we use nDCG@10 and recall@100 as the evaluation metrics. In detail, we transform the binary/3-scale judgments into 0-1/0-1-2 labels first, respectively and then calculate the metrics with pytrec\_eval~\cite{pytrec_eval}. Considering the realistic setting where the retrieval systems are often used in a zero-shot cross-domain scenario~\cite{ahmad-etal-2019-reqa,thakur2021beir}, we also adopt the zero-shot evaluation fashion. That is, the models can be trained on the training split of the MS MARCO dataset and should be tested on the test splits of MS MARCO as an in-domain evaluation setting and the remaining four datasets as an out-of-domain evaluation setting.

\section{Experiments}
In the experiments, we extend the SoTA passage retrievers (\autoref{sec:base-retrievers}) with document context with various approaches (\autoref{sec:introducing-document-context}).

\subsection{Base Retrievers}
\label{sec:base-retrievers}
We experiment with BM25 and neural passage retrievers. We use PySerini~\cite{pyserini} with the default setting for BM25 retrieval. For the neural retrievers, we use: (1)~\textbf{DRAGON+}\footnote{The checkpoint from \url{https://huggingface.co/facebook/dragon-plus-query-encoder} and \url{https://huggingface.co/facebook/dragon-plus-context-encoder}.}~\cite{lin-etal-2023-train}, (2)~\textbf{SPLADEv2}\footnote{The checkpoint from \url{https://huggingface.co/naver/splade-cocondenser-ensembledistil}.}~\cite{spladv2distil}, and (3)~\textbf{ColBERTv2}\footnote{The checkpoint from \url{https://huggingface.co/colbert-ir/colbertv2.0}.}
~\cite{santhanam-etal-2022-colbertv2}. All of them are trained on MS MARCO and they represent the main neural-retrieval architectures with both strong in-domain/out-of-domain performance. More details are available at~\autoref{sec:neural_passage_retrieval}.

\begin{table*}[t]
\centering
\resizebox{14cm}{!}{
\begin{tabular}{|L{2cm}|C{1.6cm}|C{1.6cm}|C{1.6cm}|C{1.6cm}|C{1.6cm}|C{1.6cm}||C{1.6cm}|} 
\hhline{-------||-}
\textbf{Model} & \textbf{MS}  & \textbf{NQ}  & \textbf{CQA} & \textbf{Genomics} & \textbf{MIRACL} & \textbf{Avg.} & \textbf{NQ-hard}  \\ 
\hline
\multicolumn{8}{|l|}{\textbf{Passage-only encoding w/o document context}}                                                                                   \\ 
\hline
BM25           & 44.5          & 24.1          & 12.0          & 32.8              & 28.7            & 28.4                       & 0.0$^*$               \\
DRAGON+        & 69.5          & 47.7          & 21.8          & 37.2              & 48.4            & 44.9                       & 0.0$^*$               \\
SPLADEv2       & 68.1          & 46.7          & 19.5          & 39.2              & 50.8            & 44.9                       & 0.0$^*$               \\
ColBERTv2      & 68.7          & 46.6          & 21.6          & 43.5              & 51.8            & 46.4                       & 0.0$^*$               \\ 
\hhline{=======:b:=}
\multicolumn{8}{|l|}{\textbf{Hybrid retrieval with BM25}} \\ 
\hline 
\multicolumn{8}{|l|}{\textit{Rank fusion: BM25 on documents + neural on passages}}                                                                          \\ 
\hline
DRAGON+        & 70.4          & 49.0          & 23.5          & 40.8              & 54.3            & 47.6                       & 3.5               \\
SPLADEv2       & 69.7          & 48.5          & 21.6          & 43.6              & 55.6            & 47.8                       & 2.6               \\
ColBERTv2      & 69.9          & 48.2          & 23.4          & 44.7              & \textbf{55.8}   & \textbf{48.4}              & 0.8               \\ 
\hline
\multicolumn{8}{|l|}{\textit{Hierarchical retrieval: BM25 on documents $\rightarrow$ neural on passages}}                                                   \\ 
\hline
DRAGON+        & 68.8          & 47.9          & 23.0          & 40.7              & 53.5            & 46.8                       & 2.4               \\
SPLADEv2       & 68.4          & 47.5          & 20.9          & 44.2              & 55.3            & 47.2                       & 2.0               \\
ColBERTv2      & 68.2          & 46.6          & 22.7          & \textbf{45.0}     & 54.8            & 47.5                       & 0.5               \\ 
\hhline{=======:b:=}
\multicolumn{8}{|l|}{\textbf{Contextualized passage representations }}                                                                                      \\ 
\hline
\multicolumn{8}{|l|}{\textit{Prepending titles }}                                                                                                            \\ 
\hline
DRAGON+        & \textbf{72.3} & 53.6          & \textbf{29.2} & 25.8              & 50.6            & 46.3                       & 33.6              \\
SPLADEv2       & 70.9          & 54.0          & 26.9          & 31.5              & 53.4            & 47.3                       & \textbf{39.8}     \\
ColBERTv2      & 71.4          & \textbf{54.2} & 28.1          & 31.6              & 53.9            & 47.9            & 39.2              \\ 
\hline
\multicolumn{8}{|l|}{\textit{Prepending document keyphrases}}                                                                                               \\ 
\hline
DRAGON+        & 69.9          & 49.1          & 27.6          & 34.3              & 48.5            & 45.9                       & 13.3              \\
SPLADEv2       & 68.1          & 49.2          & 26.3          & 35.4              & 50.5            & 45.9                       & 16.7              \\
ColBERTv2      & 69.6          & 48.7          & 26.6          & 41.0              & 52.8            & 47.7                       & 11.8              \\ 
\hline
\multicolumn{8}{|l|}{\textit{Coreference resolution }}                                                                                                      \\ 
\hline
DRAGON+        & 69.0          & 49.5          & 22.8          & 37.3              & 48.6            & 45.4                       & 13.0              \\
SPLADEv2       & 67.8          & 48.2          & 20.5          & 39.0              & 51.2            & 45.4                       & 12.7              \\
ColBERTv2      & 68.5          & 48.4          & 21.9          & 42.5              & 52.5            & 46.8                       & 11.8              \\
\hhline{-------||-}
\end{tabular}}
\caption{Evaluation using nDCG@10. The best result for each column is in \textbf{bold}. MS, NQ, and CQA stand for MS MARCO, Natural Questions, and ConditionalQA, respectively. Evaluation scores indicated by * are not comparable to the others, as the queries in NQ-hard are selected by making these entries all zero deliberately. Note that there is a rank fusion step at the end stage of hierarchical retrieval. Refer to~section~\ref{sec:hybrid_retrieval_approaches} for details.}
\label{tbl:main_results}
\end{table*}

\subsection{Introducing Document Context}
\label{sec:introducing-document-context}
We experimented with two ways of introducing document context to the neural passage retrievers: hybrid retrieval with BM25 and contextualized passage representations.

\subsubsection{Hybrid retrieval with BM25}
\label{sec:hybrid_retrieval_approaches}
Since the neural retrievers mainly accept short-length passages (e.g. less than 512 tokens) while BM25 does not have such a length limit, an intuitive thought is to use BM25 to retrieve whole documents and the neural retrievers to retrieve passages. We study two such hybrid-retrieval approaches:

\begin{description}[wide,itemindent=\labelsep]
    \item[Rank fusion] fuses the relevance scores from a BM25 retriever and a neural retriever. We compute the fusion as the convex combination of the normalized relevance scores~\cite{dense-retrievers-require-interpolation}:
    \begin{align*}
        &s_{\mathrm{convex}}(q, p, d) \\
    &=\alpha \hat{s}_{\mathrm{BM25}}(q, p) + (1- \alpha) \hat{s}_{\mathrm{neural}}(q, d),
    \end{align*}
    where $\alpha\in[0,1]$ is the fusion weight, and $\hat{s}_{\mathrm{BM25}}$/$\hat{s}_{\mathrm{neural}}$ represents the normalized BM25/neural-retrieval relevance score, respectively. The normalization for a relevance score $s$ is calculated as:
    \begin{equation}
        \label{eq:normalization}
        \hat{s}(q, c) = \frac{s(q, c) - m_q}{ M_q - m_q},    
    \end{equation}
    where $c$ represents the passage/document candidate, $m_q$ and $M_q$ are the min. and max. relevance scores of the top candidates for $q$, respectively. For any misaligned candidates, a zero score is taken for the candidate-missing side. In this work, the fusion is applied between the BM25 document rankings and the neural passage rankings of a certain cut-off.
    
    \item[Hierarchical retrieval] fulfills the task through two steps: (1) document retrieval and (2) passage retrieval within the retrieved documents. In~\citet{liu-etal-2021-dense-hierarchical}, the document retrieval is applied via neural retrieval on the document summaries (e.g. abstracts, table of contents, etc.) and the second step is actually a rank fusion process without normalization. For a fair comparison and a realistic setting\footnote{We do not assume that abstracts and table of contents are available.}, we apply the document retrieval step with BM25 and apply rank fusion with score normalization (cf.~\autoref{eq:normalization}) for the second step.
    
\end{description}

\subsubsection{Contextualized Passage Representations}
Although hybrid retrieval with BM25 is intuitive, it only models the document-passage interaction by combining the relevance scores rather than the actual texts. This is far from the expectation that the system can truly understand that a certain passage comes from its associated document and relies on the document context to interpret its meaning. For example, the passage in~\autoref{fig:motivative-example} might be interpreted as performance records in other music venues if it came from another document other than "The Half Moon, Putney". To fill the gap, we study three approaches to create contextualized passages. The goal is to make the passages standalone without their associated documents:
\begin{description}[wide,itemindent=\labelsep]
    \item[Prepending titles] simply adds the title text to the beginning of each passage for the same document. We use a space token\footnote{Since the base retrievers are not trained with other special tokens for this separation usage.} to separate the title text and the original passage text. The titles usually accurately show the main topic of the documents, since they are written by the author along with the document creation. One realistic concern is that such information is not always available. For example, in the MS MARCO Document Ranking task, 10.1\% of the documents from the Bing index have empty titles or nonsense titles that are composed of punctuations only.

    \item[Prepending document keyphrases] is similar to prepending titles and bypasses the title availability issue by adding the keyphrases extracted from the document instead. We use the TopicRank algorithm~\cite{bougouin-etal-2013-topicrank} with the default setting to extract the top-10 keyphrases for each document. The 10 keyphrases are concatenated by semi-colons before being prepended to the passage text with again, a space-token separator.

    \item[Coreference resolution] annotates the passages by adding the coreference information. We input the whole document into the coreference resolution model to get the mention-antecedent mappings. For each mention, its predicted antecedent shown in the earliest position\footnote{We find this heuristic performs better than others like computing TF-IDF values.} within the document is appended to it with parentheses. For example, the passage in~\autoref{fig:motivative-example} will be annotated into "Artists who have performed or recorded at the venue (The Half Moon)...". Since the in-paragraph coreference does not contribute to the document context, only the across-paragraph coreference is considered. We use the SpanBERT-large model\footnote{The checkpoint from~\url{https://storage.googleapis.com/allennlp-public-models/coref-spanbert-large-2021.03.10.tar.gz}.}~\cite{joshi-etal-2020-spanbert} fine-tuned on OntoNotes~\cite{pradhan-etal-2012-conll} with the c2f-coref approach~\cite{lee-etal-2018-higher,joshi-etal-2019-bert}. To improve memory efficiency and speed, we use distance pruning instead of the default coarse-to-fine pruning and limit the max. number of antecedent candidates for each mention to the closest 50\footnote{These modifications yield almost the same precision while increasing the acceptable input length from around 2K words to around 40K words.}. 
\end{description}

\subsection{Experiment Settings}
For both the neural retriever and the BM25 retriever, we retrieve top-1000 passages with exact search and apply the fusion over the 2000 retrieved passages if needed. We tune the fusion weight $\alpha$ on the dev split of MS MARCO. For rank fusion, the best alpha values for DRAGON+/SPLADEv2/ColBERTv2 are 0.3/0.3/0.2, respectively; for hierarchical retrieval, the best alpha values are 0.2/0.2/0.1. We carry out all the experiments in one run with 4 NVIDIA Tesla V100 GPUs. The total computational cost is around 900 GPU hours.

\section{Results}
\label{sec:results}
The main results are shown in~\autoref{tbl:main_results}.

\subsection{Hybrid Retrieval with BM25}
Both rank fusion and hierarchical retrieval can significantly improve the performance on most of the datasets by up to 5.9 nDCG@10 points on MIRACL with rank fusion and DRAGON+. Interestingly, we find rank fusion outperforms hierarchical retrieval most of the time (by up to 1.6 nDCG@10 points on MS MARCO with DRAGON+), which is not covered in the previous work. However, both approaches fail to process the hard queries in NQ-hard, achieving less than 3.5 nDCG@10 for all the performance scores. This shows that these two approaches achieve improvement mainly by enhancing the performance on the self-contained cases rather than solving the hard ones, where understanding the document context is necessary.

    \subsection{Contextualized Passage Representations}
    As a type of document summary, prepending titles and prepending document keyphrases share a similar trend: both of them can improve the performance on all the datasets except Genomics, where a significant drop (by up to 11.9 nDCG@10 point with ColBERTv2 and prepending titles) is observed. The underlying reason is discussed in detail in~\autoref{sec:why_title_harms}. Coreference resolution in general performs the worst\footnote{We also experiment with combining prepending titles and coreference resolution, but we find no further improvement.}, yielding improvement on only NQ, CQA, and MIRACL. Despite that rank fusion achieves better averaged performance than the contextualized passage representations, the latter outperforms the former on the NQ-hard, yielding up to 38.4 nDCG@10 points gain by prepending titles. This shows the importance of distinguishing the self-contained cases from the hard cases for evaluating the DAPR ability.

\begin{figure}[t]
  \centering
  \includegraphics[width=70mm]{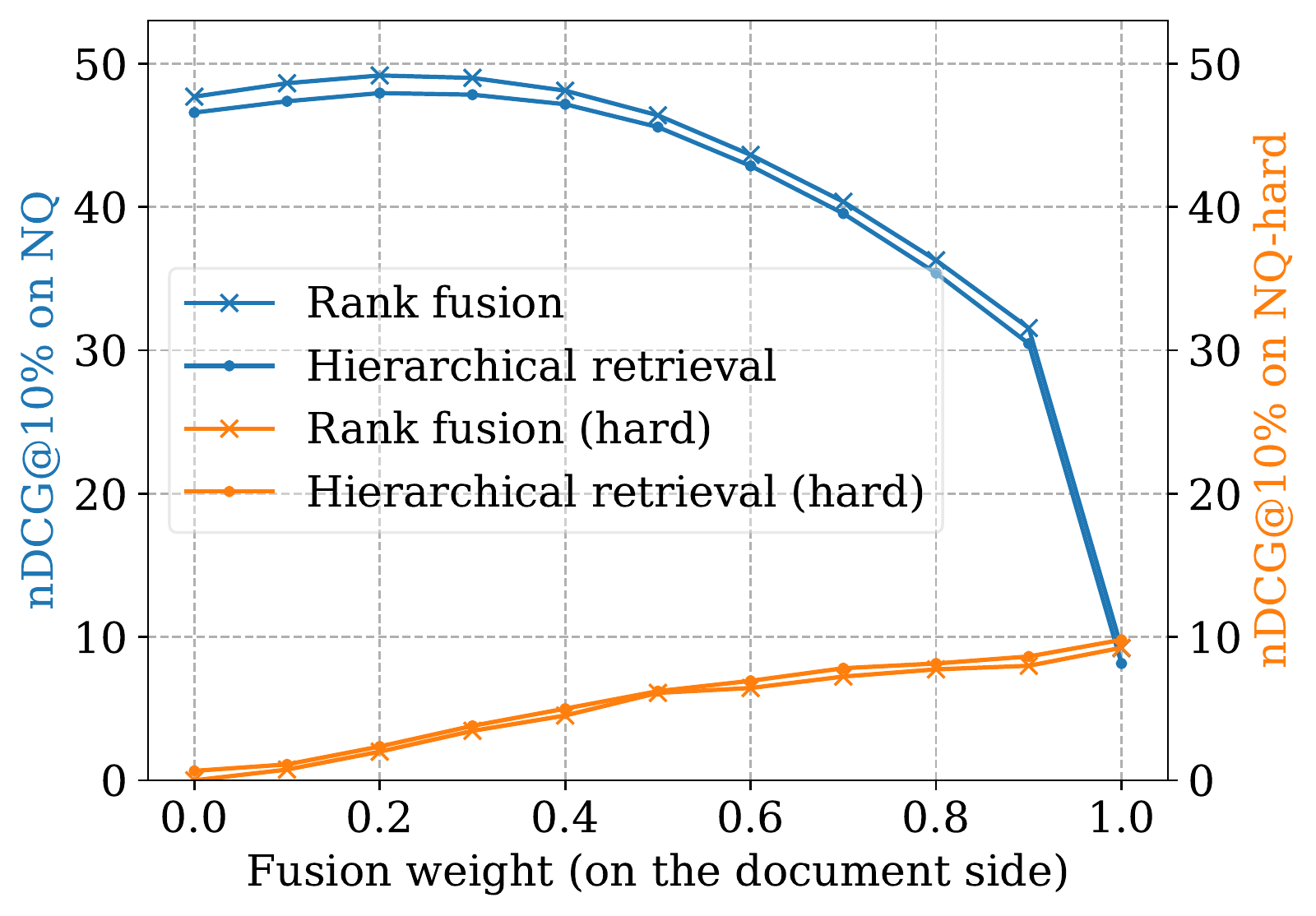}
  \caption{Influence of the fusion weight in hybrid retrieval on the passage-retrieval performance on NQ and NQ-hard. The results show that the best fusion weight on NQ cannot be directly transferred to NQ-hard.}
  \label{fig:alpha_vs_doc_retrieval}
\end{figure}

\begin{table}
\centering
\resizebox{7cm}{!}{
\begin{tabular}{|l|C{1.6cm}|C{1.6cm}|} 
\hline
\textbf{Approach}                 & \textbf{Passage} & \textbf{Document}  \\ 
\hline
Passage only (BM25)            & 0.0         & 44.3              \\
Passage only                   & 0.0         & 63.9              \\
\hline
Rank fusion                    & 3.5         & 67.9              \\
Hierarchical retrieval         & 2.4         & 67.1              \\
\hline
Prepending titles              & 33.6         & 75.9              \\
Prepending doc. keyphrases & 13.3         & 67.2              \\
Coreference resolution         & 13.0         & 67.3              \\
\hline
\end{tabular}}
\caption{Passage- vs. document-retrieval evaluation on NQ-hard using nDCG@10. The document retrieval is achieved by MaxP. The results show that rank fusion and hierarchical retrieval can retrieve documents effectively but fail to rank the passages from these documents.}
\label{tbl:doc_retrieval}
\end{table}

\section{Discussion}
In this section, we discuss the reasons behind the results and carry out an error analysis. The neural retriever is DRAGON+ by default.

\subsection{Why Does Hybrid Retrieval Perform Poorly for Hard Queries?}
In~\autoref{tbl:main_results}, compared with the contextualized passage representations, the hybrid retrieval approaches achieve comparable or better effectiveness on the full sets of the evaluation data, but they fail to process the hard queries. To understand the reason, we evaluate the systems on document retrieval by directly mapping their query-passage relevance scores equal to the query-document relevance scores for obtaining document rankings. This simple mapping approach is referred to as MaxP in~\citet{ance}. The results are shown in~\autoref{tbl:doc_retrieval}. We find the document-retrieval performance of hybrid retrieval is now comparable to that of the contextualized passage representations. This indicates hybrid retrieval does help retrieve the relevant document, but it cannot rank the relevant passages effectively from these documents.

Another issue is about the conflicts on the fusion weight between the BM25 relevance on the documents and the neural relevance on the passages (cf.~\autoref{sec:hybrid_retrieval_approaches}). We plot the dynamics of the passage-retrieval performance on NQ and NQ-hard along with different fusion weights in~\autoref{fig:alpha_vs_doc_retrieval}. We find the evaluation score on the full set of queries (i.e. NQ) peaks at a small fusion weight of 0.2 (on the document side), while that for the hard queries (i.e. NQ-hard) peaks at a very large fusion weight of 1.0. This means it is difficult for hybrid retrieval to achieve a good performance on NQ and NQ-hard at the same time.

\begin{table}
\centering
\resizebox{7.5cm}{!}{
\begin{tabular}{|l|C{1.5cm}|C{1.5cm}|C{1.5cm}|C{1.5cm}|C{1.5cm}|} 
\hline
\textbf{Context}  & \textbf{MS} & \textbf{NQ} & \textbf{CQA} & \textbf{Genomics} & \textbf{MIRACL}  \\ 
\hline
W/o titles & 9.2          & 6.8          & 10.5          & 3.4               & 6.1              \\
W/ titles & 9.4          & 7.6          & 14.9          & 3.2               & 6.3              \\ 
\hline
$\Delta$             & +0.2         & +0.8         & +4.4          & -0.2              & +0.2             \\
\hline
\end{tabular}}
\caption{Jaccard similarity between the query and the gold relevant passage when prepending titles or not. Prepending titles decreases the Jaccard similarity on Genomics while it increases the value on the other datasets.}
\label{tbl:jaccard}
\end{table}

\subsection{Why Does Prepending Titles Harm on Genomics?}
\label{sec:why_title_harms}
Despite that prepending title is viewed as a safe choice for better effectiveness in previous work~\cite{karpukhin-etal-2020-dense,thakur2021beir,liu-etal-2021-dense-hierarchical}, we find it is not held true on the Genomics dataset, where prepending titles and document keyphrases yield a significant performance drop systematically. To understand the reason, we calculate the Jaccard similarity between the query and the gold relevant passage and report the values in~\autoref{tbl:jaccard}. We find only on Genomics, prepending titles leads to lower Jaccard similarity (-0.2), while the value increases by 0.2$\sim$4.4 on the other datasets. This means the title text in Genomics usually contains many new entities or keywords that are not shown in the query or the passage, distracting the retriever from matching the query-passage pair. In addition, the much longer title length (22.0 tokens on average, cf.~\autoref{tbl:stats}) exacerbates this issue. An example is shown in~\autoref{tbl:genomics_example}. 


\begin{figure}[t]
  \centering
  \includegraphics[width=70mm]{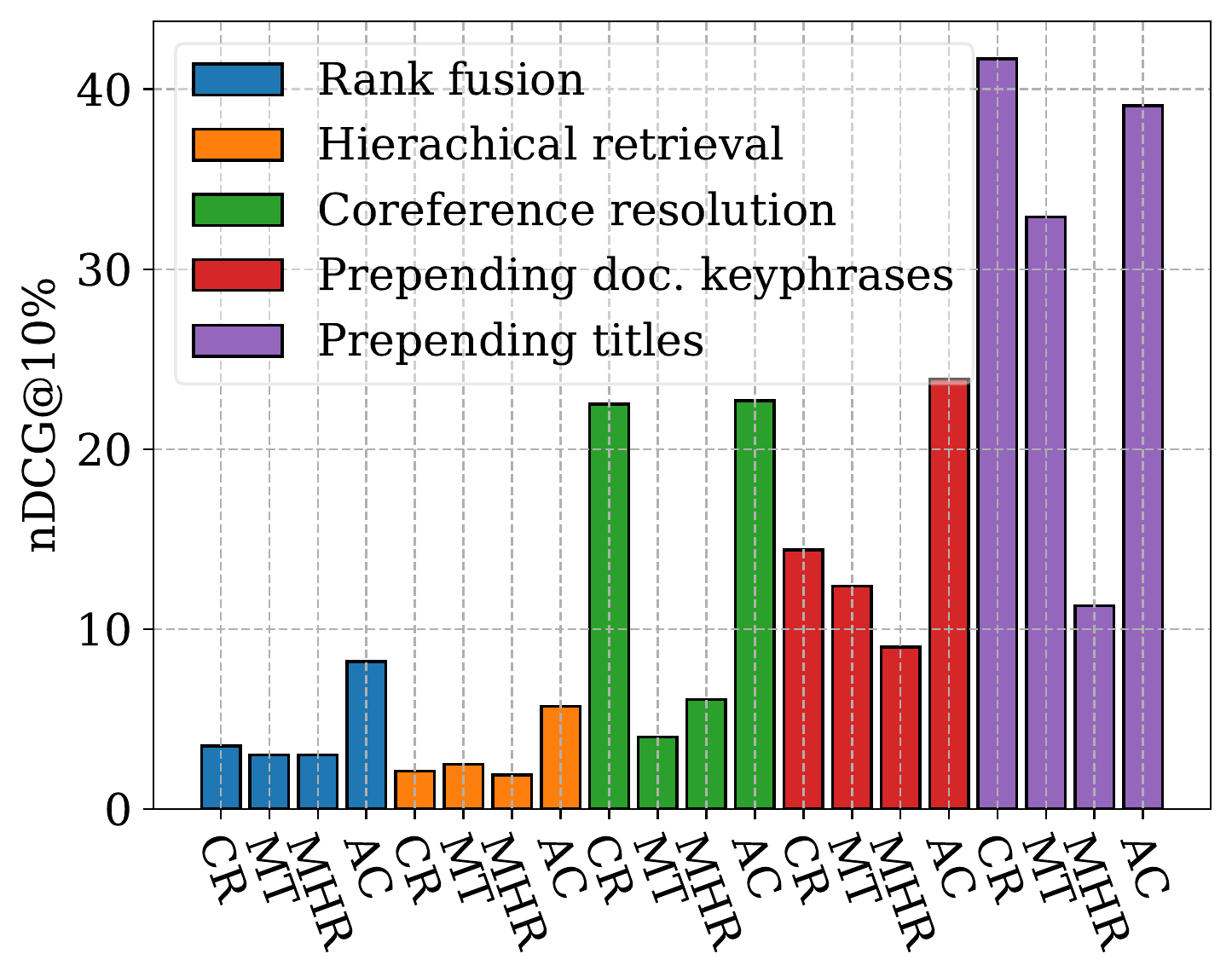}
  \caption{Evaluation scores on different categories (CR: Coreference Resolution, MT: Main Topic, MHR: Multi-Hop Reasoning, and AC: Acronym, cf.~\autoref{sec:nq_hard}) in NQ-hard. All approaches fall short on MHR and achieve good performance on AC.}
  \label{fig:categorized_performance}
\end{figure}
\subsection{Error Anlysis}
As a benefit of having NQ-hard and its categorization annotation (cf.~\autoref{tbl:nq_hard}), we can conveniently analyze the strengths and weaknesses of the retrieval systems on DAPR in detail by comparing their evaluation scores on each category of the hard cases. The results are shown in~\autoref{fig:categorized_performance}. We find multi-hop reasoning is the most difficult category, where all the approaches achieve relatively low performance (1.9$\le$nDCG@10$\le$11.3); on the other hand, acronym is the easiest category for each approach (5.7$\le$nDCG@10$\le$39.1). Interestingly,  we find that on the CR category using the CR approach underperforms simply prepending titles (by 19.2 nDCG@10), despite its significant advantage over the other approaches (by up to 20.4 nDCG@10). We suppose this is because the title usually contains the antecedent (e.g. "The Half Moon, Putney" in~\autoref{fig:motivative-example}) of the coreference mention (e.g. "the venue" in) and the CR approach can annotate the passage multiple times, disturbing term matching.

\section{Conclusion}
Inspired by the error analysis for the SoTA retrievers (DRAGON+, SPLADEv2 and ColBERTv2), we find the major errors (53.5\% in~\autoref{tbl:nq_hard}) are due to the cases where the document context is necessary. However, they cannot find the correct passages because of missing coreference resolution, the main topic of the associated document, multi-hop reasoning, etc. In these cases, the passage and the document context jointly provide the key information. This drives us to develop the DAPR benchmark, which specifically tests the retriever systems if they can take the document context into account. 

We test different ways to model such document context for passage retrieval. We find that while hybrid retrieval performs the strongest on the mixture of the easy and the hard queries, it completely fails on the hard queries that require document-context understanding. Instead, we also test informing the passage representation with document context and we can observe good improvement on these hard queries, but overall these approaches also perform rather poorly. This opens up new interesting research opportunities to develop better retrieval approaches that are capable of understanding the document context during passage retrieval. Our created benchmark allows the research community to create, test, and compare such approaches.

\section{Limitations}
NQ-hard is one dataset based on Wikipedia. Due to the annotation effort, we can only create the hard dataset for NQ. Doing it for other datasets as well is interesting. But for domain-specialized, e.g. Genomics, it requires high-skill annotation in the domain. We encourage the research community to create more datasets that require document-context understanding. Another limitation is the research is solely based on English. For other languages, it could expose other linguistic phenomena going beyond, for example, coreference in this work. Besides the narrow discussion of information retrieval and question answering, the issue of missing document context can be further studied in other tasks like fact-checking or text generation.

\section{Ethical Concerns}
All the datasets in DAPR are publicly available. We provide the processing scripts to process the original data under our format and requirements.

\section*{Acknowledgments}
This work has been funded by the German Research Foundation (DFG) as part of the QASciInf project (grant GU 798/18-3).

\bibliography{anthology,custom}
\clearpage

\appendix

\section{Dataset Details}
\label{sec:dataset_details}

\begin{description}[wide,itemindent=\labelsep]
    \item[MS MARCO] is originally a Question-Answering
    (QA) dataset built with queries from the Bing
    search log and passages from the Bing index (Nguyen et al., 2016). During its annotation,
    the annotator is provided with independent passage candidates (likely from different documents)
    retrieved by Bing search along with their source document URLs. Based on this QA task, there are also a Passage Ranking task and a Document Ranking task~\cite{trecdl19}. We combine all of these three into a single dataset to meet the DAPR requirement. In detail, we apply regex
    fuzzy match with a 5-mismatch allowance between
    the QA passage and its associated document in the
    corpus. The QA pairs whose passage span cannot
    be located in any documents are discarded (around
    50\% of the cases\footnote{The corpus is crawled after the creation of the QA dataset and the mismatch is mainly because many pages have been updated. And increasing the mismatch allowance cannot reduce this number significantly.}). We then build the gold labels by
    viewing the question/passage in each remaining QA
    pair as the query/gold-relevant passage in DAPR. Different from the other datasets in DAPR, MS MARCO does not provide the full paragraph segmentations for its documents and only a few selected paragraphs for each document are presented in its Passage Ranking task. Thus, we limit the candidate passage pool to the passages in the Passage Ranking task. The original dataset is provided under the “MIT License” for non-commercial research purposes.
    

    \item[Natural Questions] are originally a fact-seeking Question-Answering (QA) dataset built with queries from the Google search log and documents from the Wikipedia pages~\cite{kwiatkowski-etal-2019-natural}. In its annotation process, the top-5 candidate Wikipedia pages are returned by the Google search engine and the annotator is asked to select the earliest HTML bounding box containing enough information to infer the answer. In the original dataset, the answers are categorized into long/short and paragraph/table/list answers. We keep only the QA pairs with long 
    paragraph answers, since these examples are more challenging. The corpus is built by gathering all the gold-relevant passages. The original dataset is provided under the CC BY-SA 3.0 license.
    \item[MIRACL] is a multilingual information-retrieval dataset~\cite{miracl} and we use the English subset. Its corpus is composed of paragraphs from the English Wikipedia dump. Each query is written by human annotators based on the first-100 words in a randomly sampled Wikipedia page. The annotator is asked to write queries that cannot be answered by these first-100 words. The candidate passages are retrieved from Wikipedia paragraphs using an ensemble model of multiple retrievers. The annotators then annotate the top-10 candidates with binary judgments. The original dataset is provided under the Apache License 2.0 license.
    \item[Genomics] is a passage retrieval task for biomedical question answering~\cite{genomics2006,genomics2007}. We combined its TREC 2006 Genomics Track and TREC 2006 Genomics Track in DAPR. The queries are biomedical questions about biological objects (e.g. genes, proteins, etc.)/processes (e.g. physiological processes or diseases) and their explicit relationship (e.g. \textit{causes}, \textit{contributes to}, etc.). The corpus is composed of scientific articles distributed by Highwire Press\footnote{\url{https://www.highwirepress.com/}}. Expert judges are involved to annotate for each query 1000 candidate passages from a pool of submitted runs. Three-level relevance is adopted: definitely relevant, possibly relevant, and not relevant. Such relevance is defined as: in general, a passage is definitely/possibility relevant if it contains all/majority of the required elements of the question and it answers/possibly answers the question, respectively. 
    \item[ConditionalQA] is originally a task of answering questions related to UK policies given a specific scenario about the asker's condition~\cite{sun-etal-2022-conditionalqa}. Each QA instance is labeled with evidence from a UK government policy webpage\footnote{\url{https://www.gov.uk}}. We take all such webpages in the original dataset as the corpus. Each such webpage is originally parsed into HTML tags, which we view as passages. The HTML tags are removed\footnote{We find this slightly improves the effectiveness of all the systems.} and only the pure natural language is kept. For each QA instance, we concatenate the scenario and the question to form a query and take the corresponding evidences as the gold-relevant passages. The original dataset is provided under the CC BY-SA 4.0 License.
    
\end{description}

\begin{figure}[t]
  \centering
  \includegraphics[width=70mm]{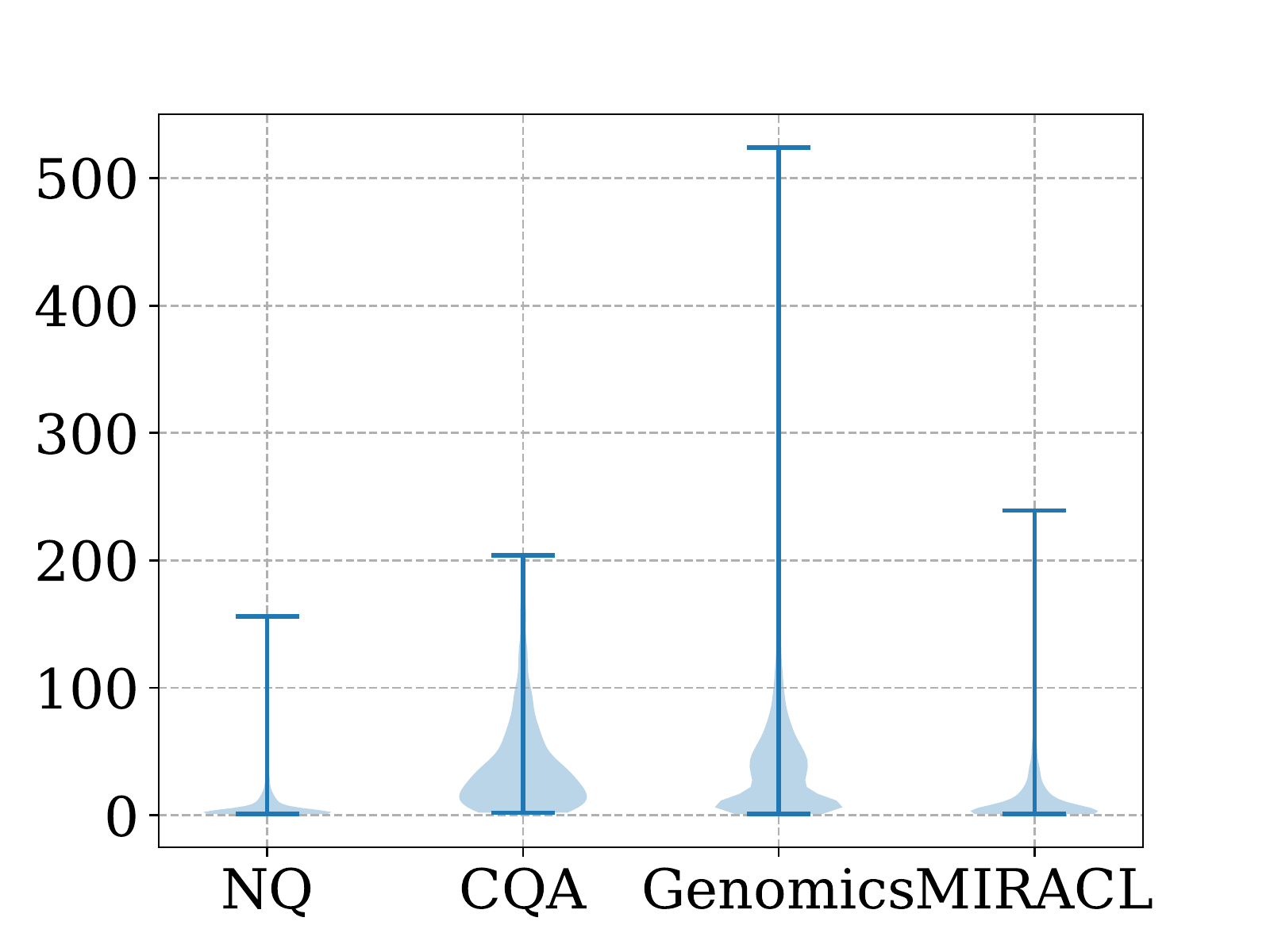}
  \caption{Distribution plots of depth of relevant passage within its associated document.}
  \label{fig:depth_dists}
\end{figure}

\section{Neural Passage Retrieval}
\label{sec:neural_passage_retrieval}
Neural passage retrieval maps queries and passages into vector representations, modeling query-passage relevance as the distance between the corresponding vectors. Single-vector dense retrieval simply embeds the input text into a single fixed-sized vector and computes cosine-similarity or dot-product between the query and the passage vectors~\cite{karpukhin-etal-2020-dense,xiao-etal-2022-retromae}. Dense retrieval is limited in modeling expressivity as its vector representation is usually low-dimensional (e.g. 768D) due to efficiency concerns. Sparse retrieval improves modeling expressivity by mapping the input text into a long sparse vector (usually vocabulary-sized) and computes dot-product as query-passage relevance~\cite{deepimpact,spladev2}. Alternatively, other work like poly-encoder~\cite{poly-encoder} and late-interaction~\cite{santhanam-etal-2022-colbertv2} represents the input text with multiple vectors and aggregates the vector distance between these multiple-vector representations. All of these neural retrievers can only accept short-length texts, e.g. 512 tokens, limiting their application scenarios. 

In this work, we use: (1) \textbf{DRAGON+}\footnote{The checkpoint from \url{https://huggingface.co/facebook/dragon-plus-query-encoder} and \url{https://huggingface.co/facebook/dragon-plus-context-encoder}.}~\cite{lin-etal-2023-train}, a dense retriever which is pre-trained with Masked Auto-Encoder on English Wikipedia, BookCorpus~\cite{BookCorpus} and the MS MARCO corpus~\cite{msmarco} and then finetuned with cross-entropy on the MS MARCO with data augmentation; (2) \textbf{SPLADEv2}\footnote{The checkpoint from \url{https://huggingface.co/naver/splade-cocondenser-ensembledistil}.}~\cite{spladv2distil}, a sparse retriever which is pre-trained with coCondensor~\cite{gao-callan-2022-unsupervised} on the pre-training corpora of RetroMAE and then finetuned with knowledge distillation on the MS MARCO training split; (3) \textbf{ColBERTv2}\footnote{The checkpoint from \url{https://huggingface.co/colbert-ir/colbertv2.0}.}
~\cite{santhanam-etal-2022-colbertv2}, a late-interaction retriever trained with knowledge distillation and cross-entropy on the MS MARCO training split. These retrievers represent the three main architectures for neural retrieval. Their training data are of a general language and do not focus on a specialized domain. All of them achieve both strong in-domain performance on the MS MARCO passage ranking task and zero-shot out-of-domain performance. In experiments, we apply exact search\footnote{I.e. brute-force search via comparing all the combinations between the queries and the passages.} over the whole corpus for all these three retrievers. We use the original settings of the pooling method and the similarity function for each retriever.

\begin{table}[t]
\centering
\resizebox{7.5cm}{!}{
\begin{tabular}{|l|C{2cm}|C{2cm}|C{2cm}|} 
\hline
\begin{tabular}[c]{@{}l@{}}\textbf{Preprocessing }\\\textbf{approach}\end{tabular}      & \begin{tabular}[c]{@{}c@{}}\textbf{Speed}\\\textbf{(word/s)}\end{tabular} & \begin{tabular}[c]{@{}c@{}}\textbf{Total}\\\textbf{latency (h)}\end{tabular} & \textbf{nDCG@10}  \\ 
\hline
Passage-only           & –                                                                         & –                                                                            & 44.9                \\
Coreference resolution & 2986.1                                                                 & 167.4                                                                          & 45.9                \\
Keyphrase extraction   & 618.8                                                                     & 819.2                                                                        & 45.4                \\
\hline
\end{tabular}}
\caption{Preprocessing costs. For coreference resolution, the speed and the latency are calculated based on 1 GPU; for keyphrase extraction, the numbers are based on 1 CPU process. The nDCG@10 numbers are copied from~\autoref{tbl:main_results}.}
\label{tbl:cost}
\end{table}

\section{Preprocessing Cost}
Besides effectiveness, the preprocessing costs are also very important, as the retrieval corpus can usually be as large as millions of documents or billions of tokens. We apply coreference resolution with 4 NVIDIA Tesla V100 GPUs and apply keyphrase extraction with 32 processes of Intel(R) Xeon(R) Platinum 8168 CPU @ 2.70GHz. The processing costs on MS MARCO are shown in~\autoref{tbl:cost}. Doing keyphrase extraction is very slow on CPU, taking 25.6 hours with 32 CPU processes (corresponding to 819.2 hours per CPU process).

\begin{table*}[t]
\centering
\resizebox{14cm}{!}{
\begin{tabular}{|L{2cm}|C{1.6cm}|C{1.6cm}|C{1.6cm}|C{1.6cm}|C{1.6cm}|C{1.6cm}||C{1.6cm}|} 
\hhline{-------||-}
\textbf{Model} & \textbf{MS}  & \textbf{NQ}  & \textbf{CQA} & \textbf{Genomics} & \textbf{MIRACL} & \textbf{Avg.} & \textbf{NQ-hard}  \\ 
\hline
\multicolumn{8}{|l|}{\textbf{Passage-only encoding w/o document context}}                                                                                   \\ 
\hline
BM25           & 44.5          & 24.1          & 12.0          & 32.8              & 28.7            & 28.4                       & 0.0$^*$               \\
Jinav2$^\dagger$	&64.6	&50.4	&25.7	&32.6	&42.2	&43.1	& 4.0 \\
DRAGON+        & 69.5          & 47.7          & 21.8          & 37.2              & 48.4            & 44.9                       & 0.0$^*$               \\
SPLADEv2       & 68.1          & 46.7          & 19.5          & 39.2              & 50.8            & 44.9                       & 0.0$^*$               \\
ColBERTv2      & 68.7          & 46.6          & 21.6          & 43.5              & 51.8            & 46.4                       & 0.0$^*$               \\ 
\hhline{=======:b:=}
\multicolumn{8}{|l|}{\textbf{Hybrid retrieval with BM25}} \\ 
\hline 
\multicolumn{8}{|l|}{\textit{Rank fusion: BM25 on documents + neural on passages}}                                                                          \\ 
\hline
DRAGON+        & 70.4          & 49.0          & 23.5          & 40.8              & 54.3            & 47.6                       & 3.5               \\
SPLADEv2       & 69.7          & 48.5          & 21.6          & 43.6              & 55.6            & 47.8                       & 2.6               \\
ColBERTv2      & 69.9          & 48.2          & 23.4          & 44.7              & {55.8}   & {48.4}              & 0.8               \\ 
\hline
\multicolumn{8}{|l|}{\textit{Rank fusion: JinaV2$^\dagger$ on documents + neural on passages}}                                                                          \\ 
\hline
DRAGON+ &71.1 &53.6 &22.1 &35.7 &49.7 &46.4 &6.1 \\
SPLADEv2 &70.1 &53.4 &20.5 &39.6 &52.5 &47.2 &4.9 \\
ColBERTv2 &70.9 &51.9 &22.3 &44.5 &53.8 &48.7 &0.0 \\
\hhline{-------||-}
\end{tabular}}
\caption{Evaluation using nDCG@10. The results except with JinaV2 are copied from~\autoref{tbl:main_results}. $\dagger$ indicates the model is trained on Natural Questions. MS, NQ, and CQA stand for MS MARCO, Natural Questions, and ConditionalQA, respectively. Evaluation scores indicated by * are not comparable to the others, as the queries in NQ-hard are selected by making these entries all zero deliberately.}
\label{tbl:long_doc_retrievers}
\end{table*}

\section{Long-Document Retriever}
In~\autoref{sec:results}, we use BM25 as the document retriever. We also experiment with a SoTA neural long-document retriever, JinaV2~\cite{jinav2}, whose input length is 8192 tokens. The results are shown in~\autoref{tbl:long_doc_retrievers}. When it is used in the rank fusion approach, we find it cannot outperform BM25 significantly except on Natural Questions, which is included in the training data of JinaV2. On NQ-hard, it also performs poorly, achieving nDCG@10 below 10 for all the cases.

\begin{table*}[t]
\centering
\resizebox{14cm}{!}{
\small
\begin{tabular}{|m{2cm}|m{3cm}|m{10cm}|} 
\hline
\textbf{Dataset}                                           & \textbf{Query}                                                                                                                                                                & \textbf{Relevant passage}                                                                                                                                                                                                                                                                                                                                                                                                                                                                                                                                                                                                                                                                                                                                                                                                                                                                                                                                                                                                                                                                                                                                                                                                                                                                                                                                                                                                                                                                                                                                \\ 
\hline
MS MARCO                                                   & which molecule is split apart in photosystem ii?                                                                                                                              & \begin{tabular}[c]{@{}m{10cm}@{}}\textit{Title:~Does Photosystem-I split water?}\\\textit{Question Asked a year ago Fisseha Asmelash9.78 Ethiopian Biodiversity Institute Does Photosystem-I split water?the chlorophyll complexes are surrounded by water. we know PS-I and PS-II both absorb photons to ...}\\\textbf{Water is split by a OEC (oxygen evolving complex) containing ions of Mn, Cl, calcium and oxygen atoms present associated with PS II; not with PS I. Also electrons required to neutralise charge created by exit of electrons from the reaction centre chlorophyll molecules are first donated by tyrosine residues present in the PS II component.}\\\textit{So PS I is not designed to split water.1 Recommendationa year ago David W. Lawlor Formerly Rothamsted Research Hello,As said well by others, PSII and associated components are responsible. Thanks to ...}\end{tabular}                                                                                                                                                                                                                                                                                                                                                                                                                                                                                                                                                                                                                                                 \\ 
\hline
\begin{tabular}[c]{@{}m{2cm}l@{}}Natural\\Questions\end{tabular} & is season 7 of homeland the last season                                                                                                                                       & \begin{tabular}[c]{@{}m{10cm}@{}}\textit{Title:~Homeland (season 7)}\\\textit{The seventh season of the American television drama series Homeland premiered on February 11, 2018, on Showtime, and will consist of 12 episodes.}\\...\\\textbf{The series was renewed for a seventh and eighth season in August 2016.[21] For this season, Maury Sterling, Jake Weber and Linus Roache were promoted to series regulars; Sterling has been recurring since the first season, while Weber and Roache both first appeared in the sixth season.[2] The seventh season began production on September 11, 2017, filming in Richmond, Virginia.[22][23] Filming wrapped in Budapest, Hungary on March 29, 2018.[24]}\\\textit{...}\end{tabular}                                                                                                                                                                                                                                                                                                                                                                                                                                                                                                                                                                                                                                                                                                                                                                                                                      \\ 
\hline
ConditionalQA                                              & I'm 71, and am currently living in rented accommodation with my 64-year-old Civil Partner. I have an existing Housing Benefit claim. Can I continue to claim Housing Benefit? & 
\begin{tabular}[c]{@{}m{10cm}@{}}\textit{Title: Housing Benefit}\\\textit{\textless{}h1\textgreater{}Eligibility\textless{}h1\textgreater{}}\\\textit{\textless{}p\textgreater{}Housing Benefit can help you pay your rent if you're unemployed, on a low income or claiming benefits. It's being replaced by Universal Credit.\textless{}p\textgreater{}}\\\textit{...}\\\textbf{\textless{}p\textgreater{}Your existing claim will not be affected if, before 15 May 2019, you:\textless{}p\textgreater{}}\\\textbf{\textless{}li\textgreater{}were getting Housing Benefit\textless{}li\textgreater{}}\\\textbf{\textless{}li\textgreater{}had reached State Pension age\textless{}li\textgreater{}}\\\textbf{\textless{}p\textgreater{}It does not matter if your partner is under State Pension age.\textless{}p\textgreater{}}\\\textit{...}\end{tabular}

\\ 
\hline

Genomics                                                   & How do interactions between insulin-like GFs and the insulin receptor affect skin biology?                                                                                    & \begin{tabular}[c]{@{}m{10cm}@{}}\textit{Title: The Regulation of Skin Proliferation and Differentiation in the IR Null Mouse: Implications for Skin Complications of Diabetes}\\\textit{Abstrat Impaired wound healing of skin is one of the most serious complications of diabetes. However, the pathogenesis of this process is not known, and it is unclear whether impaired insulin signaling could directly affect skin physiology. To elucidate the role of insulin in skin, we studied skin insulin receptor (IR) null mice. The morphology of the skin of newborn IR null mice was normal; however, these mice exhibited decreased proliferation of skin keratinocytes and changes in expression of skin differentiation markers ...}\\...\\\textbf{Differences in the differentiation of IR-KO keratinocytes in vivo vs. that of cells cultured and induced to differentiate in vitro were also found. The differentiation process of intact skin seemed to be slightly impaired, whereas it is significantly altered in cultured keratinocytes of the same animal. The main change observed in the IR-KO skin was an increase in K6 expression in the IR-KO epidermis compared with the IR-WT skin. In normal murine and human epidermis, K6 is expressed constitutively in a variety of internal stratified epithelia as well as in palmoplantar epidermis and specialized cells of the hair follicle (20). K6 expression was also shown to be induced during wound healing, skin disorders such as psoriasis, skin tumors ...}\\...\end{tabular}  \\ 
\hline
MIRACL                                                     & How large is the Caraga Administrative Region?                                                                                                                                & \begin{tabular}[c]{@{}m{10cm}@{}}\textit{Title: Caraga}\\\textit{Caraga, officially known as the Caraga Administrative Region or simply Caraga Region and designated as Region XIII, is an administrative region in ...}\\...\\\textbf{The region has a total land area of, representing 6.3\% of the country's total land area and 18.5\% of the island of Mindanao. 47.6\% of the total land area of the region belongs to the province of Agusan del Sur. Of the total land area, 71.22\% is forestland and 28.78\% is alienable and disposable land. Major land uses include forestland comprising 31.36\% and 23.98\% of agricultural and open spaces.}\\...\end{tabular}                                                                                                                                                                                                                                                                                                                                                                                                                                                                                                                                                                                                                                                                                                                                                                                                                                                                                 \\
\hline
\end{tabular}}
\caption{Examples of queries and relevant passages in DAPR. Gold-relevant passages are \textbf{bold} and other document context is \textit{italic}. The HTML tags (e.g. <p>, <li>, etc.) in ConditionalQA are removed in preprocessed data.}
\label{tbl:examples}
\end{table*}

\begin{table*}[t]
\centering
\resizebox{16cm}{!}{
\begin{tabular}{|L{2cm}|C{1.6cm}|C{1.6cm}|C{1.6cm}|C{1.6cm}|C{1.6cm}|C{1.6cm}||C{1.6cm}|} 
\hhline{-------||-}
\textbf{Model} & \textbf{MS}  & \textbf{NQ}  & \textbf{CQA} & \textbf{Genomics} & \textbf{MIRACL} & \textbf{Avg.} & \textbf{NQ-hard}  \\ 
\hline
\multicolumn{8}{|l|}{\textbf{Passage-only encoding w/o document context}}                                                                                   \\ 
\hline
BM25           &80.8 &63.3 &38.5 &19.5 &71.6 &54.8 &12.2$^*$               \\
DRAGON+        &96.9 &85.3 &48.0 &19.7 &85.7 &67.1 &39.7$^*$              \\
SPLADEv2       &96.2 &83.8 &47.0 &23.5 &87.3 &67.6 &31.5$^*$               \\
ColBERTv2      &95.5 &82.5 &41.9 &23.6 &88.0 &66.3 &24.5$^*$              \\ 
\hhline{=======:b:=}
\multicolumn{8}{|l|}{\textbf{Hybrid retrieval with BM25}} \\ 
\hline 
\multicolumn{8}{|l|}{\textit{Rank fusion: BM25 on documents + neural on passages}}                                                                          \\ 
\hline
DRAGON+        &97.5 &89.3 &56.2 &25.8 &90.4 &71.8 &56.2               \\
SPLADEv2       &96.5 &87.8 &55.1 &27.9 &90.8 &71.6 &48.1               \\
ColBERTv2      &96.2 &84.8 &46.4 &25.0 &90.3 &68.5 &34.2              \\ 
\hline
\multicolumn{8}{|l|}{\textit{Hierarchical retrieval: BM25 on documents $\rightarrow$ neural on passages}}                                                   \\ 
\hline
DRAGON+        &92.7 &87.0 &54.3 &27.2 &88.4 &69.9 &60.0               \\
SPLADEv2       &92.4 &85.2 &52.5 &\textbf{29.1} &89.2 &69.7 &49.7              \\
ColBERTv2      &92.1 &81.9 &43.8 &25.6 &88.3 &66.3 &32.7               \\ 
\hhline{=======:b:=}
\multicolumn{8}{|l|}{\textbf{Contextualized passage representations }}                                                                                      \\ 
\hline
\multicolumn{8}{|l|}{\textit{Prepending titles }}                                                                                                            \\ 
\hline
DRAGON+        &\textbf{98.2} &\textbf{93.9} &\textbf{70.1} &16.6 &89.3 &73.6 &\textbf{86.1}             \\
SPLADEv2       &97.5 &93.2 &67.6 &21.5 &90.2 &\textbf{74.0} &84.6    \\
ColBERTv2      &97.1 &92.8 &64.8 &18.3 &\textbf{91.4} &72.9 &83.2             \\ 
\hline
\multicolumn{8}{|l|}{\textit{Prepending document keyphrases}}                                                                                               \\ 
\hline
DRAGON+        &97.4 &90.6 &69.0 &19.9 &86.9 &72.8 &67.1             \\
SPLADEv2       &96.4 &89.6 &65.8 &23.8 &88.8 &72.9 &64.3              \\
ColBERTv2      &96.3 &88.9 &62.5 &24.0 &89.8 &72.3 &61.8              \\ 
\hline
\multicolumn{8}{|l|}{\textit{Coreference resolution }}                                                                                                      \\ 
\hline
DRAGON+        &96.9 &87.3 &48.2 &19.9 &86.4 &67.8 &52.5             \\
SPLADEv2       &96.2 &86.6 &47.0 &23.2 &87.4 &68.1 &48.4             \\
ColBERTv2      &95.6 &85.3 &42.3 &23.3 &88.6 &67.0 &43.4             \\
\hhline{-------||-}
\end{tabular}}
\caption{Evaluation using recall@100. The best result for each column is in \textbf{bold}. MS, NQ, and CQA stand for MS MARCO, Natural Questions, and ConditionalQA, respectively. The corresponding recall@100 results are available in~\autoref{tbl:main_results_recall}. * indicates the evaluation scores are not comparable to the others, as the queries in NQ-hard are selected by making the corresponding nDCG@10 all zero deliberately. For more details, please refer to~\autoref{sec:nq_hard}.}
\label{tbl:main_results_recall}
\end{table*}

\begin{table}
\centering
\resizebox{7.5cm}{!}{
\begin{tabular}{|L{2cm}|L{2.2cm}|L{4.8cm}|} 
\hline
\textbf{Query}                                                                                                                                                                     & \textbf{Ttile}                                                                                                                                                                                                                                                                        & \textbf{Passage}          \\ 
\hline
What is the role of \textcolor{red}{APC} (\textcolor{red}{\uline{adenomatous}} \textcolor{red}{\uline{polyposis}} \textcolor{red}{\uline{coli}}) in colon \textcolor{red}{cancer}? &  The \textcolor{blue}{Ubiquitin}-\textcolor{blue}{Proteasome} \textcolor{blue}{Pathway} and \textcolor{blue}{Serine} \textcolor{blue}{Kinase} \textcolor{blue}{Activity}\textcolor{blue}{Modulate} \uline{Adenomatous} \uline{Polyposis} \uline{Coli} Protein-\textcolor{blue}{mediated} \textcolor{blue}{Regulation} of~\textcolor[rgb]{0.129,0.129,0.129}{$\beta$}\textcolor[rgb]{0.129,0.129,0.129}{-}Catenin-\textcolor{blue}{Lymphocyte} \textcolor{blue}{Enhancer-binding} \textcolor{blue}{Factor} \textcolor{blue}{Signaling}* & Mutations in the tumor suppressor \textcolor{red}{\uline{adenomatous}} \textcolor{red}{\uline{polyposis}} \textcolor{red}{\uline{coli}}(\textcolor{red}{APC})1 gene are responsiblefor tumors that arise in both familial \textcolor{red}{\uline{adenomatous}} \textcolor{red}{\uline{polyposis}} andsporadic \textcolor{red}{colon} \textcolor{red}{cancer}s (1-7). \textcolor{red}{APC} mutations are almostalways truncating, giving rise to proteins lacking C termini (6, 8, 9).Efforts to understand how these mutations contribute to \textcolor{red}{cancer} havefocused on the ability of \textcolor{red}{APC} to bind and subsequently down-regulatethe cytoplasmic levels of $\beta$-catenin (10-13).  \\
\hline
\end{tabular}}
\caption{An example corresponding to a decreased Jaccard similarity (from 10.9 to 7.9) when prepending title on Genomics. The words in \textcolor{red}{red} are the common words between the query and the passage; the \uline{underlined} words are the common words among the query, the title, and the passage; the words in \textcolor{blue}{blue} are the new words not shown in the query and the passage. Prepending titles introduces many new entities/concepts and distracts the query-passage matching.}
\label{tbl:genomics_example}
\end{table}

\end{document}